\documentclass[useAMS,usenatbib]{mn2e}
\usepackage{mn2e-breakabs}
\usepackage{graphicx}
\usepackage{times}
\newcommand{\Hunit}{\,{\rm km}\,{\rm s}^{-1}\,{\rm Mpc}^{-1}}

\def\fun#1#2{\lower3.6pt\vbox{\baselineskip0pt\lineskip.9pt
        \ialign{$\mathsurround=0pt#1\hfill##\hfil$\crcr#2\crcr\sim\crcr}}}

\def\bfk{\mbox{\bf k}}

\def\bfs{\mbox{\bf s}}

\def\rmd{\mbox{d}}

\def\rd{{\rm d}}

\newcommand{\be}{\begin{equation}}
\newcommand{\ee}{\end{equation}}
\newcommand{\ba}{\begin{eqnarray}}
\newcommand{\ea}{\end{eqnarray}}
\newcommand{\simgt}{\,\hbox{\lower0.6ex\hbox{$\sim$}\llap{\raise0.6ex\hbox{$>$}}}\,}
\newcommand{\simlt}{\,\hbox{\lower0.6ex\hbox{$\sim$}\llap{\raise0.6ex\hbox{$<$}}}\,}

\begin{document}

\title[$H(z)$, $D_A(z)$, and $f_g(z)\sigma_8(z)$ from SDSS DR9 BOSS Data]
{Model-Independent Measurements of Cosmic Expansion and Growth at $z=0.57$
Using the Anisotropic Clustering of CMASS Galaxies
From the Sloan Digital Sky Survey Data Release 9}

\author[Yun Wang]{
  \parbox{\textwidth}{
    Yun Wang\thanks{E-mail: wang@nhn.ou.edu}}
  \vspace*{4pt} \\
  Homer L. Dodge Department of Physics \& Astronomy, Univ. of Oklahoma,
                 440 W Brooks St., Norman, OK 73019, U.S.A.\\
                 }

\date{\today}

\maketitle

\begin{abstract}

We analyze the anisotropic two dimensional galaxy correlation function (2DCF) of the CMASS galaxy samples
from the Sloan Digital Sky Survey Data Release 9 (DR9) of the Baryon Oscillation Spectroscopic Survey 
(BOSS) data. Modeling the 2DCF fully including nonlinear effects and redshift space distortions (RSD)
in the scale range of 30 to 120 $h^{-1}$Mpc,
we find $H(0.57)r_s(z_d)/c=0.0444 \pm 0.0019$, $D_A(0.57)/r_s(z_d)=9.01 \pm 0.23$,
and $f_g(0.57)\sigma_8(0.57)=0.474 \pm 0.075$,
where $r_s(z_d)$ is the sound horizon at the drag epoch computed using a simple integral,
and $f_g(z)$ is the growth rate at redshift $z$, and $\sigma_8(z)$ represents the matter
power spectrum normalization on $8\,h^{-1}$Mpc scale at $z$.
We find that the scales larger than 120 $h^{-1}$Mpc are dominated by noise in the 2DCF
analysis, and that the inclusion of scales 30-40 $h^{-1}$Mpc significantly tightens
the RSD measurement. Our measurements are consistent with previous results using
the same data, but have significantly better precision since we are using all the
information from the 2DCF in the scale range of 30 to 120 $h^{-1}$Mpc.
Our measurements have been marginalized over sufficiently wide priors for
the relevant parameters; they can be combined with other data to probe dark energy and gravity.

\end{abstract}


\section{Introduction}  \label{sec:intro}

Galaxy clustering (GC) is one of the most powerful probes in our continuing quest to illuminate
the mystery of cosmic acceleration \citep{Riess98,Perl99}, and differentiate between its two possibles causes:
an unknown energy component in the Universe (i.e., dark energy), or modification of
general relativity (i.e., modified gravity).\footnote{For recent reviews, see 
\cite{Ratra08,Frieman08,Caldwell09,Uzan10,Wang10,Li11,Weinberg12}.}
This is because galaxy clustering enables the measurement of cosmic expansion history \citep{Blake03,Seo03},
as well as the growth history of cosmic large scale structure \citep{Guzzo08,Wang08}.
At present, our largest GC data set comes from the Baryon Oscillation Spectroscopic Survey (BOSS) 
[part of the Sloan Digital Sky Survey (SDSS) III], which will obtain galaxy redshifts over
10,000 square degrees up to a redshift of 0.7 upon completion in 2014 \footnote{http://www.sdss3.org/surveys/boss.php}.
The Euclid space mission, scheduled for launch in 2020, will obtain galaxy redshifts over 15,000 square degrees
over a wide redshift range up to a redshift of two \footnote{http://www.euclid-ec.org/}\citep{RB}.

The SDSS Data Release 9 (DR9) provides us with the first public data set for galaxy clustering from BOSS.
In this paper, we build on methods first presented in \cite{CW12}, \cite{CW13}, \cite{WCH13}, and 
\cite{Hemantha13}, and present an independent new analysis of the DR9 BOSS
galaxy clustering data. The main differences between this analysis and previous work using
the same data are: (1) We utilize all the available information (not just the multipoles) in the
anisotropic two dimensional galaxy correlation function (2DCF) of the CMASS galaxy samples
of DR9 BOSS data, and obtain model-independent constraints on the cosmic expansion rate
$H(z)$, the angular-diameter distance $D_A(z)$, and the normalized growth rate $f_g(z)\sigma_8(z)$
\citep{Song09} (with $f_g(z)$ denoting the growth rate at redshift $z$, and $\sigma_8(z)$ denoting the matter
power spectrum normalization on $8\,h^{-1}$Mpc scale at $z$). 
(2) We marginalize over sufficiently wide priors for
$\Omega_m h^2$, $\Omega_b h^2$, $n_s$, $P_0$, as well as parameters used
to model nonlinear effects and RSD; thus our results 
can be combined with other data to probe dark energy and gravity.

We present our method in Section~\ref{sec:method}, our results in Section~\ref{sec:results}, and
summarize and conclude in Section~\ref{sec:conclusion}.

\section{Methodology}
\label{sec:method}

\subsection{Modeling the Galaxy Correlation Function}
\label{subsec:model}

We model the two point galaxy correlation function by convolving the 2DCF with linear RSD with
a distribution for galaxy peculiar velocities $f(v)$:
\be
\label{eq:model}
 \xi(\sigma,\pi)=\int_{-\infty}^\infty \tilde{\xi}\left(\sigma,\pi-\frac{v}{H(z)a(z)}
 \right)\,f(v)dv,
\ee
where $H(z)$ is the Hubble parameter and $a(z)$ is the cosmic scale factor,
and $f(v)$ is given by \citep{Ratcliffe98,Landy02}
\be
 f(v)=\frac{1}{\sigma_v\sqrt{2}}\exp\left(-\frac{\sqrt{2}|v|}{\sigma_v}\right),
\ee
with $\sigma_v$ denoting the galaxy peculiar velocity dispersion.

The 2DCF $\tilde{\xi}$ is the Fourier transform of the dewiggled galaxy power spectrum
\citep{Hemantha13}:
\be
P(\bfk)_{dw,nl}^{g,s}=b^2\left(1+\beta \mu^2\right)^2 F_{NL}(k)\,P_{dw,lin}(\bfk),
\ee
where $b$ is galaxy bias, $\beta$ is the linear redshift distortion parameter,
and $\mu$ is the cosine of the angle between $\bfk$ and the line-of-sight.
The linear dewiggled power spectrum is given by
\be
P_{dw,lin}(\bfk)= G^2(z)P_0 k^{n_s} \left\{T^2_{\rm nw}(k) + T^2_{\rm BAO}(k) e^{-g_\mu k^2 /(2k_*^2)}\right\},
\ee
where we have defined
\be
T^2_{\rm BAO}(k)=T^2(k)-T^2_{\rm nw}(k),
\ee
with $T(k)$ denoting the linear matter transfer function, and
$T_{\rm nw}(k)$ denoting the pure CDM (no baryons) transfer function given by
Eq.(29) from \cite{EH98}. 
The nonlinear damping factor, $e^{-g_\mu k^2 /(2k_*^2)}$, was derived by \cite{Eisen07} using N-body simulations.
The factor $g_\mu$ describes the enhanced damping along the line of sight due to the enhanced power:
\be
g_\mu(\bfk,z) \equiv G^2(z) \{ 1-\mu^2 +\mu^2 [1+f_g(z)]^2 \}.
\label{eq:gmu}
\ee
\noindent
Note that $g_\mu$ scales with the linear growth factor $G(z)$ squared, which corresponds to the
scale of the linear regime increasing with $1/G(z)$ at high redshifts.
As density perturbations grow with cosmic time, the linear regime expands
as we go to higher redshifts.
The function $F_{NL}(k)$ models nonlinear evolution and scale-dependent bias \citep{Cole05}:
\be
F_{NL}(k)=\frac{1+Qk^2}{1+Ak+Bk^2}.
\ee
We take $B=Q/10$ \citep{Sanchez08}.

In taking the Fourier transform of $P(\bfk)_{dw,nl}^{g,s}$, it is useful to write
\ba
\label{eq:Pk}
&&P(\bfk)_{dw,nl}^{g,s}=P(\bfk)_{nw,nl}^{g,s}+P(\bfk)_{BAO,dw,nl}^{g,s}\\
&&P(\bfk)_{nw,nl}^{g,s}=b^2 G^2(z)\left(1+\beta \mu^2\right)^2 P_0 k^{n_s} T^2_{\rm nw}(k) F_{NL}(k)\nonumber\\
&&P(\bfk)_{BAO,dw,nl}^{g,s}=b^2 G^2(z)\left(1+\beta \mu^2\right)^2
P_0 k^{n_s} T^2_{\rm BAO}(k)\nonumber\\
& & \hskip 3cm F_{NL}(k) e^{-g_\mu k^2 /(2k_*^2)}\nonumber.
\ea
Fourier transform of Eq.(\ref{eq:Pk}) gives
\be
\tilde{\xi}(\sigma,\pi)=\xi^{g,s}_{nw}(\sigma,\pi)+\xi_{BAO,dw}^{g,s}(\sigma,\pi),
\ee
where $\sigma$ and $\pi$ are the transverse and line-of-sight separations of a pair of galaxies, respectively.
It is most efficient to Fourier transform the two term in Eq.(\ref{eq:Pk}) separately,
as they have different dependence on $\mu$. 

The Fourier transform of $P(\bfk)_{nw,nl}^{g,s}$ is given by \citep{Hamilton92}
\ba
 \label{eq:xi_nw}
\xi^{g,s}_{nw}(\sigma,\pi)&=&b^2 G^2(z) P_0 (\xi_0^{nw}(s)P_0(\mu)+\xi_2^{nw}(s)P_2(\mu)\nonumber\\
& & \hskip 1cm +\xi_4^{nw}(s)P_4(\mu)),
\ea
where $s=\sqrt{\sigma^2+\pi^2}$, 
$\mu$ is the cosine of the angle between $\bfs=(\sigma,\pi)$ and the line-of-sight, and 
$P_l$ are Legendre polynomials. The multipoles of $\xi^{nw}$ are defined as
\begin{eqnarray} \label{eq:xi_mp}
 \xi_0^{nw}(r)&=&\left(1+\frac{2\beta}{3}+\frac{\beta^2}{5}\right)\xi^{nw}(r),\\
 \xi_2^{nw}(r)&=&\left(\frac{4\beta}{3}+\frac{4\beta^2}{7}\right)[\xi^{nw}(r)-\bar{\xi}^{nw}(r)],\\
 \xi_4^{nw}(r)&=&\frac{8\beta^2}{35}\left[\xi^{nw}(r)+\frac{5}{2}\bar{\xi}^{nw}(r)
 -\frac{7}{2}\overline{\overline{\xi}}^{nw}(r)\right],
\end{eqnarray}
where $\beta$ is the linear RSD parameter and
\begin{eqnarray} \label{eq:xi_bar}
 \bar{\xi}^{nw}(r)&=&\frac{3}{r^3}\int_0^r\xi^{nw}(r')r'^2dr',\\
 \overline{\overline{\xi}}^{nw}(r)&=&\frac{5}{r^5}\int_0^r\xi^{nw}(r')r'^4dr', \label{eq:xi_barbar}
\end{eqnarray}
where $\xi^{nw}(r)$ is given by
\be
\xi^{nw}(r)=\frac{1}{2\pi^2 r}\int_0^{\infty}\rmd k \, k^{n_s+1} T^2_{\rm nw}(k) F_{NL}(k) \sin(kr).
\ee

The Fourier transform of $P(\bfk)_{BAO,dw,nl}^{g,s}$ is more complicated due to
the additional damping factor $e^{-g_\mu k^2 /(2k_*^2)}$, where $g_\mu$ depends on $\mu$
(see Eq.[\ref{eq:gmu}]). This $\mu$-dependent damping factor in $k$-space 
becomes a Gaussian convolution in configuration space \citep{CW13}:
\be
 \label{eq:xi_gBAO}
 \xi^{g,s}_{BAO,dw}(\sigma,\pi)=\frac{1}{\sigma_\star\sqrt{\pi}}
\int_{-\infty}^\infty \rmd x\,\xi^{g,s}_{BAO,sdw}(\sigma,\pi-x)\, e^{-x^2/\sigma_\star^2},
\ee
where $\xi^{g,s}_{BAO,sdw}(\sigma,\pi)$ is the Fourier transform of $P(\bfk)_{BAO,dw,nl}^{g,s}$
with the damping factor $e^{-g_\mu k^2 /(2k_*^2)}$ replaced by its $\mu$-independent part,
$e^{-G^2(z) k^2 /(2k_*^2)}$, and 
\be
\sigma_\star^2=\frac{[4f_g(z)+2f^2_g(z)]G^2(z)}{k_\star^2}.
\ee
$\xi^{g,s}_{BAO,sdw}(\sigma,\pi)$ can be obtained using Eq. (\ref{eq:xi_nw})-(\ref{eq:xi_barbar}),
with the superscript ``nw'' replaced by ``BAO'', and $\xi^{nw}(r)$ replaced by
\be
\xi^{BAO}(r)=\frac{1}{2\pi^2 r}\int_0^{\infty}\rmd k \, k^{n_s} T^2_{\rm BAO}(k) F_{NL}(k)
e^{-G^2(z) k^2 /(2k_*^2)}
\ee

\subsection{Data and Covariance Matrix}
\label{subsec:data}

We use the CMASS samples (both North and South) of BOSS from SDSS DR9 made publicly available by the
BOSS Collaboration. The CMASS North sample consists of 207,246 galaxies,
while the CMASS South sample consists of 57,037 galaxies.
The total effective area (accounting for all applied cuts and the completeness in every sector included)
of the North and South samples is 3275 (deg)$^2$ \citep{Anderson12}. 

We convert the measured redshifts of galaxies to comoving distances 
assuming the same fiducial model as that of the mock catalogs:
$\Lambda$CDM model with $\Omega_k=0$, $h=0.7$, $\Omega_m h^2=0.13426$ ($\Omega_m=0.274$), 
$\Omega_b h^2=0.0224$ ($\Omega_b=0.0457$), $n_s=0.95$, and $\sigma_8=0.8$.
We use the \cite{Landy93} two-point correlation function estimator given by 
\begin{equation}
\label{eq:xi_Landy}
\xi(\sigma,\pi) = \frac{DD(\sigma,\pi)-2DR(\sigma,\pi)+RR(\sigma,\pi)}{RR(\sigma,\pi)},
\end{equation}
where $\pi$ is the separation along the line of sight (LOS), $\sigma$ 
is the separation in the plane of the sky, DD, DR, and RR represent the normalized 
data-data, data-random, and random-random pair counts respectively in a given
distance range. The LOS is defined as the direction from the observer to the 
center of a pair. The bin size we use here is
$10 \, h^{-1}$Mpc$\times 10 \,h^{-1}$Mpc. 
The Landy and Szalay estimator has minimal variance for a Poisson
process. We use the random data sets that accompany the BOSS data sets;
these have been generated with the same radial and angular selection functions as the real data. 
Note that the BOSS catalogs include weights that should be applied to
each galaxy.

We use the publicly available BOSS DR9 mock catalogs by \cite{Manera13},
to estimate the covariance matrix of the observed correlation function. 
We calculate the 2D correlation functions 
of the 600 mock catalogs and construct the covariance matrix as
\begin{equation}
 C_{ij}=\frac{1}{N-1}\sum^N_{k=1}(\bar{\xi}_i-\xi_i^k)(\bar{\xi}_j-\xi_j^k),
\label{eq:covmat}
\end{equation}
where $N$ is the number of the mock catalogs, $\bar{\xi}_m$ is the
mean of the $m^{th}$ bin of the mock catalog correlation functions, and
$\xi_m^k$ is the value in the $m^{th}$ bin of the $k^{th}$ mock
catalog correlation function.

\subsection{The Likelihood Analysis}

We perform a Markov Chain Monte-Carlo likelihood analysis \citep{Lewis02}
in obtaining our results. For Gaussian distributed measurements, the likelihood of 
a model given the data is proportional to $\exp(-\chi^2/2)$ \citep{Press92},
where $\chi^2$ compares data with model predictions.
For our analysis, $\chi^2$ is given by
\be
 \label{eq:chi2}
 \chi^2\equiv\sum_{i,j=1}^{N_{bins}}\left[\xi_{th}(\bfs_i)-\xi_{obs}(\bfs_i)\right]
 C_{ij}^{-1}
 \left[\xi_{th}(\bfs_j)-\xi_{obs}(\bfs_j)\right]
\ee
where $\xi_{th}$ (described in Sec.\ref{subsec:model}) and $\xi_{obs}$ 
(described in Sec.\ref{subsec:data}) are the theoretical and observed correlation functions
respectively. $N_{bins}$ is the number of data bins used, and $\bfs_i=(\sigma_i,\pi_i)$. 

Naively, $\xi_{obs}$ should be measured from the observed galaxy redshifts
and positions for each $\xi_{th}$ tested. However, the observed
galaxy distribution occupies different physical volumes in different
cosmological models. This means that the number of data bins varies from
model to model, which renders the $\chi^2$ values ill-defined in
a galaxy correlation function analysis. This problem is solved by
noting that the fiducial model is only used in converting redshifts into 
distances for the galaxies in our data sample. This means that
assuming different models in converting redshifts into distances 
results in observed galaxy distributions that are related by
simple scaling of the galaxy separations. 

The separations of galaxies in angle and redshift are observables, and
independent of the model assumed, i.e.,
\ba
&&\Delta\theta = \frac{\sigma}{D_A(z)}=\frac{\sigma_{fid}}{D_A^{fid}(z)}\\
&& \Delta z= H(z) \pi = H^{fid}(z) \pi_{fid},
\ea
where $(\sigma,\pi)$ and $(\sigma_{fid},\pi_{fid})$ are the transverse and 
line-of-sight separations of galaxies in an arbitrary model and the 
fiducial model, respectively. $\{H(z), D_A(z)\}$ and $\{H^{fid}(z), D_A^{fid}(z)\}$
are the Hubble parameter and the angular diameter distance in an arbitrary model 
and the fiducial model, respectively.
Therefore, for a thin redshift shell, we can convert 
the separation of one pair of galaxies from the fiducial model to 
another model by performing the scaling
(see, e.g., \cite{Seo03})
\be
\label{eq:scaling}
 (\sigma,\pi)=\left(\frac{D_A(z)}{D_A^{fid}(z)}\sigma_{fid},
 \frac{H^{fid}(z)}{H(z)}\pi_{fid}\right).
\ee
Consequently, the measured 2D correlation functions assuming an arbitrary model 
and the fiducial model are related as follows:
\be
\xi_{obs}(\sigma,\pi)= T\left(\xi^{fid}_{obs}(\sigma_{fid},\pi_{fid})\right)
\ee
where $T$ denotes the mapping given by Eq.(\ref{eq:scaling}).

Now we can rewrite the $\chi^2$ from Eq.(\ref{eq:chi2}) as \citep{CW12}
\ba 
\label{eq:chi2_2}
\chi^2 &\equiv&\sum_{i,j=1}^{N_{bins}}
 \left\{T^{-1}\left[\xi_{th}(\bfs_i)\right]-\xi^{fid}_{obs}(\bfs_i)\right\}
 C_{fid,ij}^{-1} \cdot \nonumber\\
 & & \cdot \left\{T^{-1}\left[\xi_{th}(\bfs_j)\right]-\xi_{obs}^{fid}(\bfs_j)\right\},
\ea
where $C_{fid}$ is the covariance matrix of the observed data 
assuming the fiducial model, and $T^{-1}\left[\xi_{th}(\bfs_i)\right]$
maps the model computed at $\{\sigma, \pi\}$ to the fiducial
model frame coordinates $(\sigma_{fid},\pi_{fid})$ as given by
Eq.(\ref{eq:scaling}).

In practice, $\xi_{th}(\sigma,\pi)$ is computed on a grid of $\{\sigma, \pi\}$,
assuming an arbitrary cosmological model parametrized by
$\Omega_m h^2$, $\Omega_b h^2$, $n_s$, $P_0$, as well as parameters used
to model nonlinear effects and RSD. This model is assumed to have 
Hubble parameter $H(z)$ and the angular diameter distance $D_A(z)$.
This means that we are using the shape of the galaxy 2PCF as
a standard ruler, with cosmological parameters ($\Omega_m h^2$, $\Omega_b h^2$, $n_s$, $P_0$)
and parameters that describe systematic effects (nonlinearity and RSD) included as calibration parameters.

To compare with data, the model is scaled to match the data grid
$\{\sigma_{fid}, \pi_{fid}\}$ using Eq.(\ref{eq:scaling}). The measured
$\xi^{fid}_{obs}(\sigma_i^{fid},\pi_i^{fid})$ is compared with the
model at 
\ba
&&\sigma_{fid} = \frac{D_A^{fid}(z)}{D_A(z)}\sigma\\
&& \pi_{fid}=\frac{H(z)}{H^{fid}(z)}\pi,
\ea
with the model multiplied by a volume factor given by
\be
V_{fac}=\frac{H(z)}{H^{fid}(z)}\left(\frac{D_A^{fid}(z)}{D_A(z)}\right)^2.
\ee

\section{Results}
\label{sec:results}

We perform a Markov Chain Monte-Carlo likelihood analysis \citep{Lewis02}.
The parameter space that we explore spans the parameter set of
$\{H(0.57), D_A(0.57), \beta, \Omega_mh^2, \Omega_bh^2, n_s, P_{norm}, \sigma_v, k_\star\, f_g(0.57), Q, A\}$,
where the dimensionless normalization parameter $P_{norm}=P_0 b^2(0.57) G^2(0.57) [h\,$Mpc$^{-1}]^{n_s+3}$.
From these parameters, we can derive the constraints on three parameters which
are well constrained and insensitive to systematic effects:
\ba
&& x_h \equiv H(0.57) r_s(z_d)/c\\
&& x_d \equiv D_A(0.57)/r_s(z_d)\\
&& f_g(0.57) \sigma_8(0.57) = I_0^{1/2} P_{norm}^{1/2} \beta,
\ea
where we have defined
\be
I_0 \equiv  \int_0^\infty\rd \bar{k}\,  \frac{\bar{k}^{n_s+2}}{2\pi^2}\, 
T^2(\bar{k}\cdot h\mbox{Mpc}^{-1})\,
\left[\frac{3 j_1(8\bar{k})}{8\bar{k}}\right]^2,
\ee
where $\bar{k}\equiv k/[h\,\mbox{Mpc}^{-1}]$, 
and $j_1(kr)$ is spherical Bessel function. Note that since $k_{\parallel}$
and $k_\perp$ scale as $H(z)$ and $1/D_A(z)$ respectively, 
the measured 2DCF does not depend on $h$ \citep{WCH13}.
However, there is an explicit $h$-dependence via the use of
$\sigma_8$, since $\sigma_8 \propto 
I_0=I_0(\omega_m, \omega_b, n_s, h)$; we compute $I_0$
with $h=0.7$ as assumed for the fiducial model.
The comoving sound horizon at redshift $z_d$ is given by
\ba
\label{eq:rs}
r_s(z_d)  &= & \int_0^{t} \frac{c_s\, dt'}{a}
=cH_0^{-1}\int_{z}^{\infty} dz'\,
\frac{c_s}{E(z')}, \\
 &= & cH_0^{-1} \int_0^{a} 
\frac{da'}{\sqrt{ 3(1+ \overline{R_b}\,a')\, {a'}^4 E^2(z')}}\nonumber\\
&=& \frac{2997.9\,\mbox{Mpc}}{\sqrt{0.75 \overline{R_b}\omega_m}}\,
\ln\left\{ \frac{\sqrt{a_d+a_{eq}}+\sqrt{a_d+\overline{R_b}^{-1}}}
{\sqrt{a_{eq}}+\sqrt{\overline{R_b}^{-1}}}
\right\}, \nonumber
\ea
where $a$ is the cosmic scale factor, $a =1/(1+z)$, and
$a^4 E^2(z)=\Omega_m (a+a_{\rm eq})+\Omega_k a^2 +\Omega_X X(z) a^4$,
with $a_{\rm eq}=\Omega_{\rm rad}/\Omega_m=1/(1+z_{\rm eq})$, and
$z_{\rm eq}=2.5\times 10^4 \Omega_m h^2 (T_{CMB}/2.7\,{\rm K})^{-4}$.
The sound speed is $c_s=1/\sqrt{3(1+\overline{R_b}\,a)}$,
with $\overline{R_b}\,a=3\rho_b/(4\rho_\gamma)$,
$\overline{R_b}=31500\Omega_bh^2(T_{CMB}/2.7\,{\rm K})^{-4}$.
We take $T_{CMB}=2.72548$ \citep{Fixsen09}.

The redshift of the drag epoch $z_d$ is given by \cite{EH98}
\begin{equation}
z_d  =
 \frac{1291(\Omega_mh^2)^{0.251}}{1+0.659(\Omega_mh^2)^{0.828}}
\left[1+b_1(\Omega_bh^2)^{b2}\right],
\label{eq:zd}
\end{equation}
where
\begin{eqnarray}
  b_1 &= &0.313(\Omega_mh^2)^{-0.419}\left[1+0.607(\Omega_mh^2)^{0.674}\right],\\
  b_2 &= &0.238(\Omega_mh^2)^{0.223}.
\end{eqnarray}

We have chosen to compute $r_s(z_d)$ using the simple formulae given above.
For a given cosmological model, this conventional choice gives a $r_s(z_d)$ value 
that differs from that given by CAMB by a factor that is close to one and 
nearly independent of the cosmological model \citep{Mehta12}.
Note that $r_s(z_d)$ is only used to scale $H(z)$ and $D_A(z)$.
As long as we use the same formulae to compute $r_s(z_d)$ in making
model predictions, the comparison between the measured and predicted values of
$\{H(0.57) r_s(z_d)/c, D_A(0.57)/r_s(z_d), f_g(0.57) \sigma_8(0.57)\}$
should be insensitive to the choice of $r_s(z_d)$.

We apply flat priors on all the parameters. The priors on $H(0.57)$, $D_A(0.57)$, $\beta$,
$\Omega_m h^2$, and $P_{norm}$ are sufficiently wide that further increasing the width of the priors 
have no effect on the results. The priors we impose on $\Omega_bh^2$
and $n_s$ are $(0.02018, 0.02438)$ and $(0.9137, 1.0187)$,
corresponding to the 7$\sigma$ range of these parameters from the first year
Planck data, with $\sigma$ from the Gaussian fits by \cite{Wang13};
these priors are wide enough to ensure that CMB constraints are not double counted 
when our results are combined with CMB data \citep{CWH12}.
Our results are not sensitive to the parameters that describe the systematic
uncertainties, $\{\sigma_v, k_\star, f_g(0.57), Q, A\}$. We have applied reasonable
flat priors on these: $\sigma_v=0-500$km/s, $k_\star/G(0.57)=0.1-0.3h/$Mpc,
$f_g(0.57)=0.35-0.55$, $Q=0-40\,($Mpc$/h)^2$, and $A=0-10\,$Mpc$/h$.

\subsection{Validation of our methodology}

\begin{figure}
\centering
\vspace{-1.0in}
\includegraphics[width=0.83\columnwidth,clip]{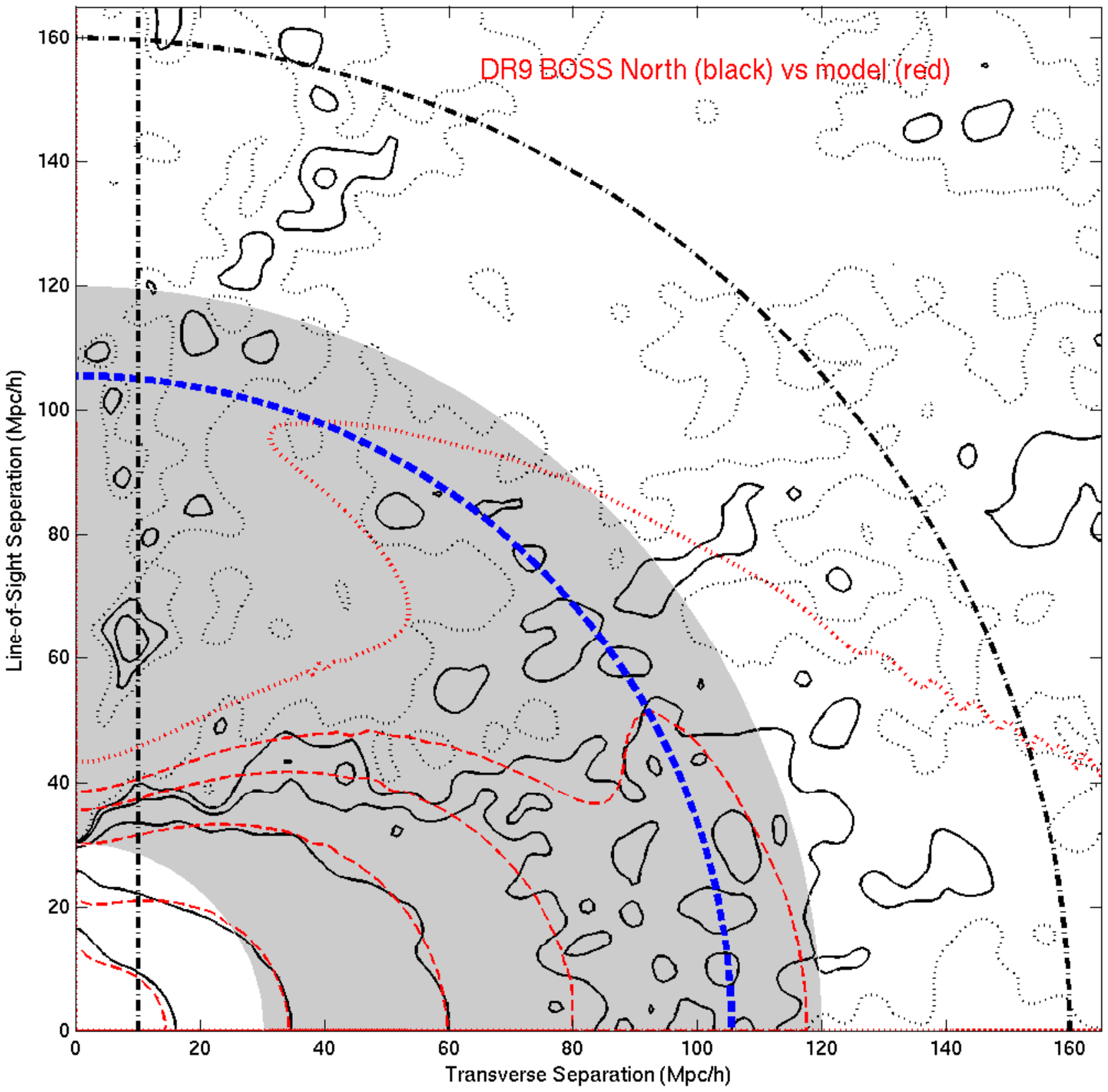}
\vspace{-1in}
\includegraphics[width=0.83\columnwidth,clip]{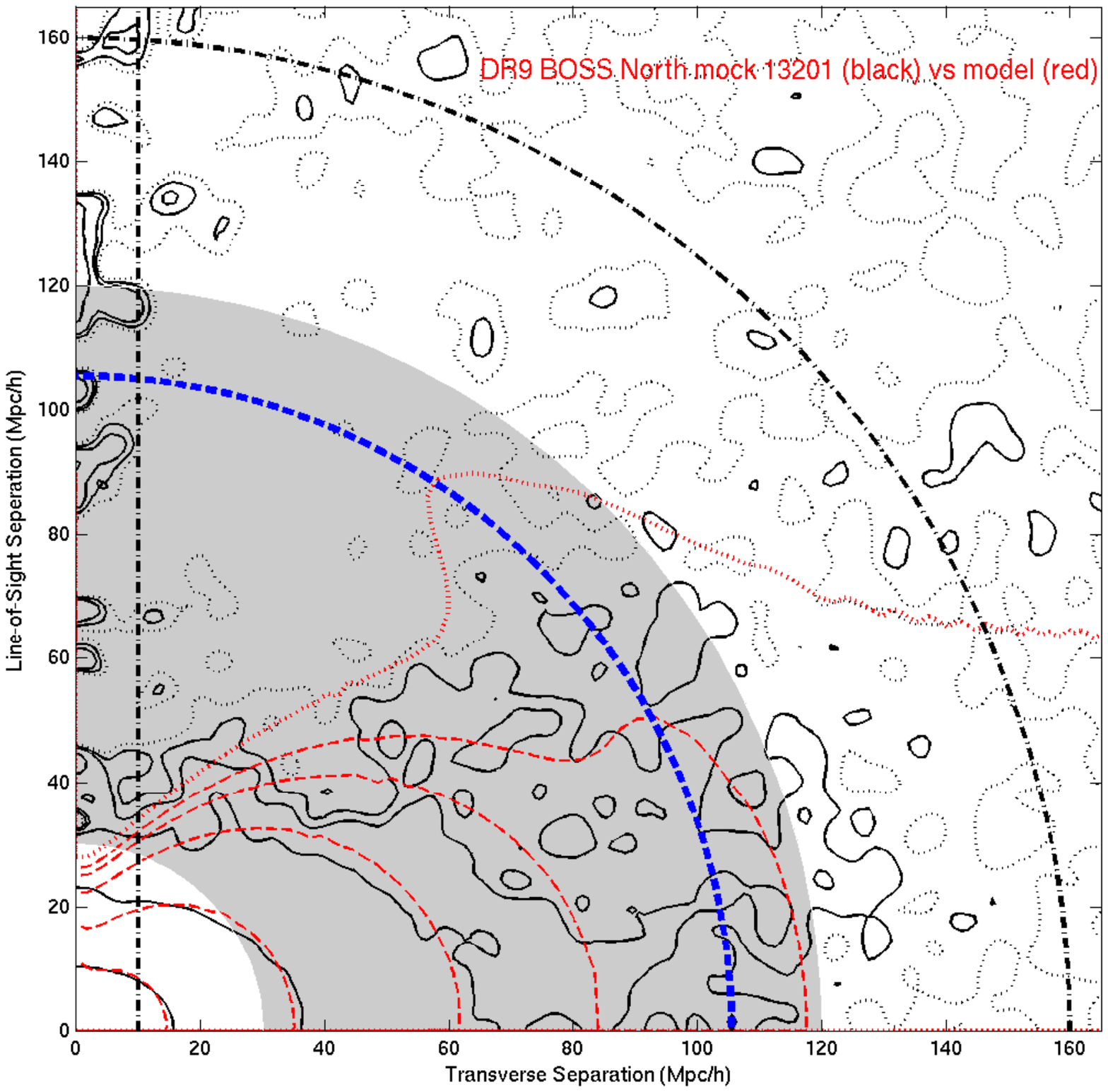}
\includegraphics[width=0.83\columnwidth,clip]{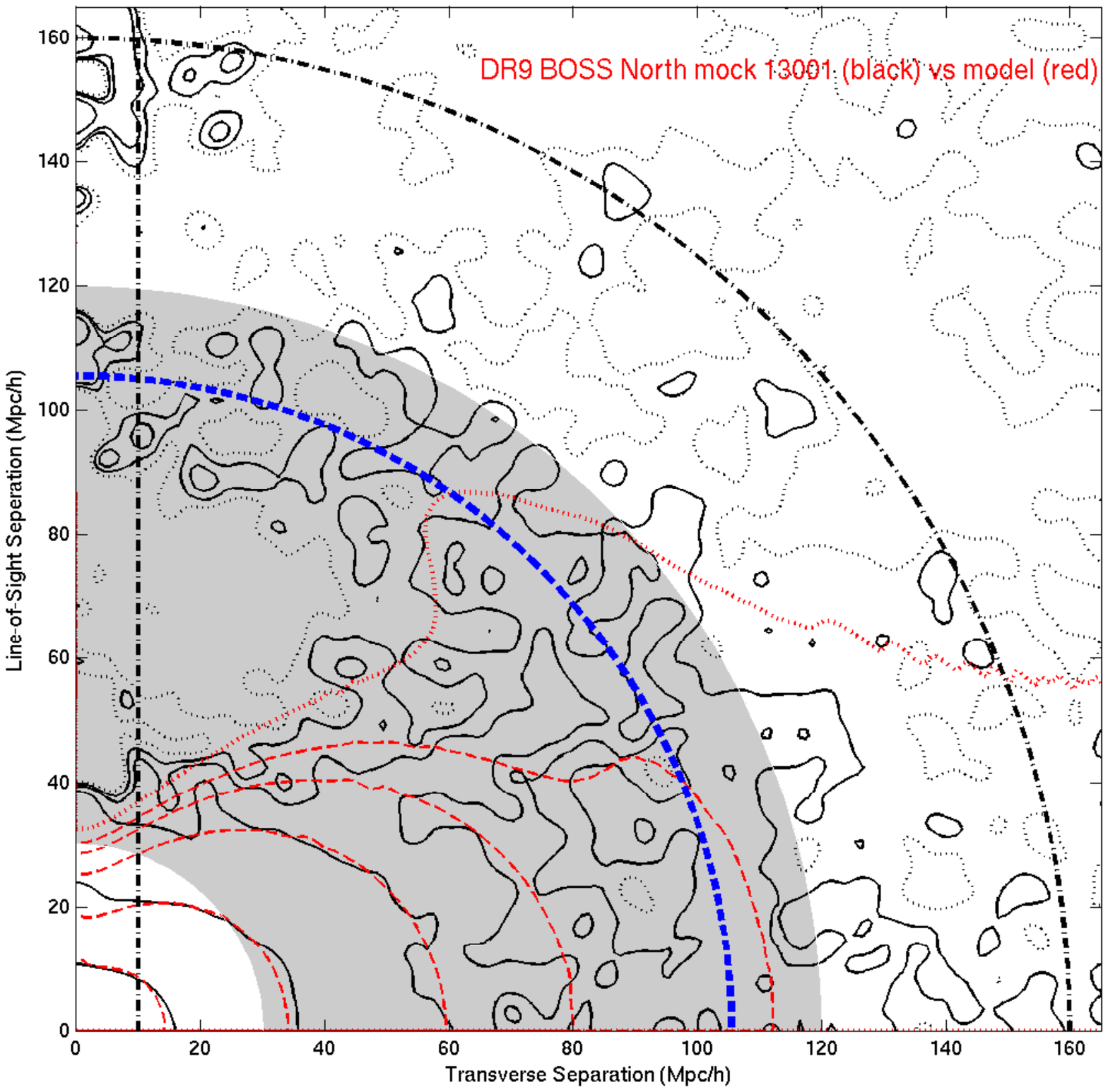}
\caption{The DR9 BOSS North CMASS sample (upper panel) and two representative mocks (middle and lower panels).
The contour levels are $\xi=0.005, 0.01, 0.025, 0.1, 0.5$, and the dotted contours denote
$\xi \leq 0$. The solid lines are the data (or mock data), and the dashed lines 
are our model (with parameters chosen from within 68\% C.L. marginalized intervals).}
\label{fig:xi2d}
\end{figure}
Figure \ref{fig:xi2d} shows the DR9 BOSS North CMASS sample (upper panel) and two mocks 
(middle and lower panels).
The contour levels are $\xi=0.005, 0.01, 0.025, 0.1, 0.5$, and the dotted contours denote
$\xi \leq 0$. The solid lines are the data (or mock data), and the dashed lines 
are our model (with parameters chosen from within 68\% C.L. marginalized intervals).
The mock in the middle panel has been chosen because it resembles the data,
but it is somewhat noisier than the data on small scales,
The mock in the lower panel is the first mock from the suite of mocks;
it is noisier than the data on all scales. 

Fig.{\ref{fig:meanxhxd} shows the distribution of $x_h=H(0.57) \,r_s(z_d)/c$ and $x_d=D_A(0.57)/r_s(z_d)$ from 104 mocks 
of the DR9 BOSS North and South CMASS samples. The solid and dashed lines denote
the likelihood peaks and marginalized means respectively. The dotted lines indicate
the values predicted by the true model of the mocks (the fiducial model assumed for the analysis of actual data).
The true values of $x_h$ and $x_d$ are within the central 68.3\% range of the recovered values. 
This validates out methodology.

\begin{figure}
\centering
\includegraphics[width=\columnwidth,clip]{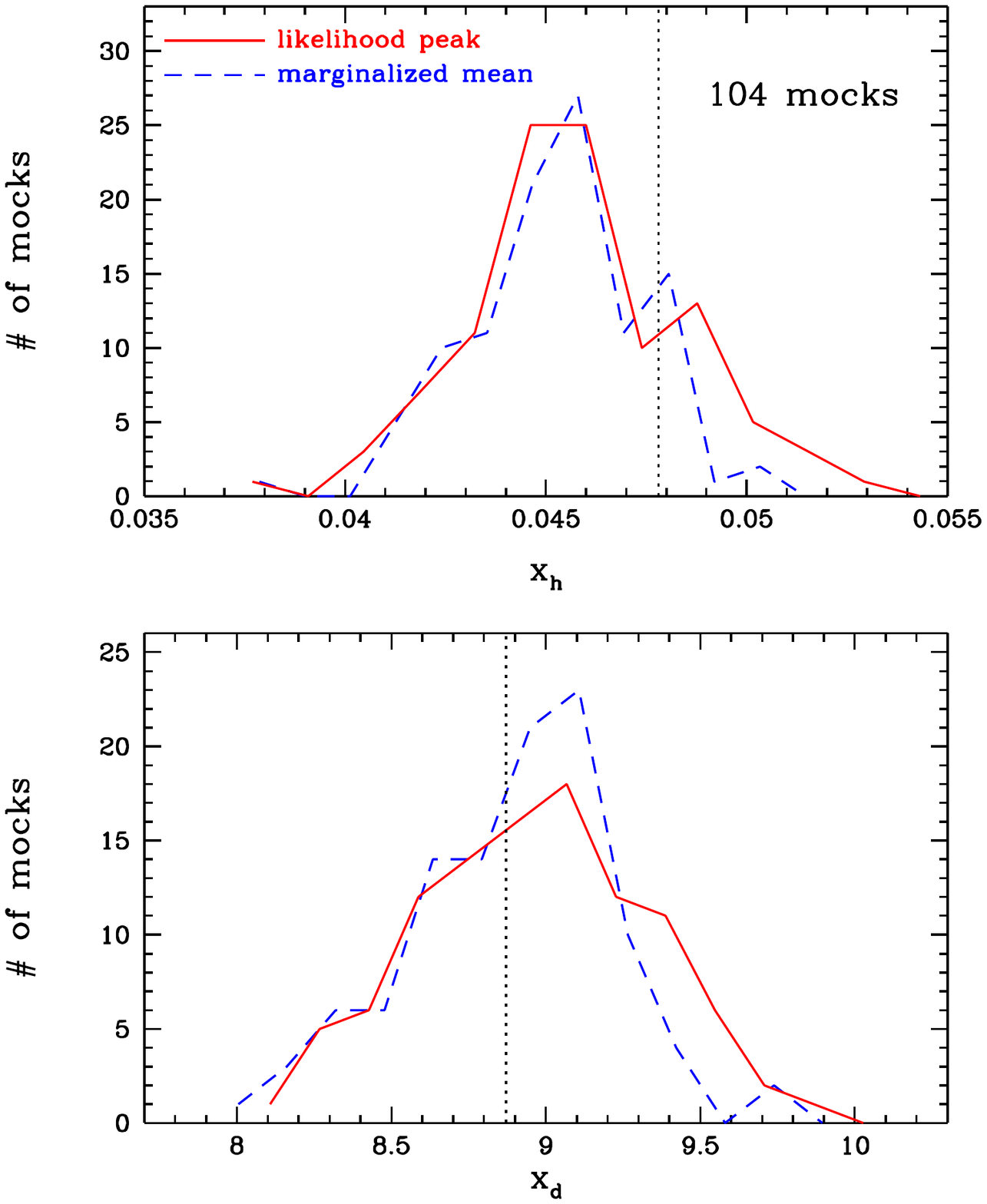}
\caption{The distribution of $x_h=H(0.57) \,r_s(z_d)/c$ and $x_d=D_A(0.57)/r_s(z_d)$ from 104 mocks 
of the DR9 BOSS North and South CMASS samples. The solid and dashed lines denote
the likelihood peaks and marginalized means respectively. The dotted lines indicate
the values predicted by the true model of the mocks (the fiducial model assumed for the analysis of actual data).}
\label{fig:meanxhxd}
\end{figure}

\subsection{Results from BOSS DR9 CMASA Samples}

We now present our results from analyzing the real data.
Fig.\ref{fig:params_pdf} shows the 1D marginalized probability distribution of parameters 
estimated from DR9 BOSS North and South CMASS samples. The different line
types denote different scale ranges used in our analysis:
30-120$\,h^{-1}$Mpc (thick solid); 30-160$\,h^{-1}$Mpc (thick dotted);
40-120$\,h^{-1}$Mpc (thin solid); 40-160$\,h^{-1}$Mpc (thin dotted).
Fig.\ref{fig:params_2D} shows the corresponding 2D joint confidence contours 
(68\% and 95\%) of the parameters, with the same line
types as in Fig.\ref{fig:params_pdf}. Only the key parameters and
parameters with significant correlations are shown in Fig.\ref{fig:params_2D}.
\begin{figure}
\centering
\includegraphics[width=\columnwidth,clip]{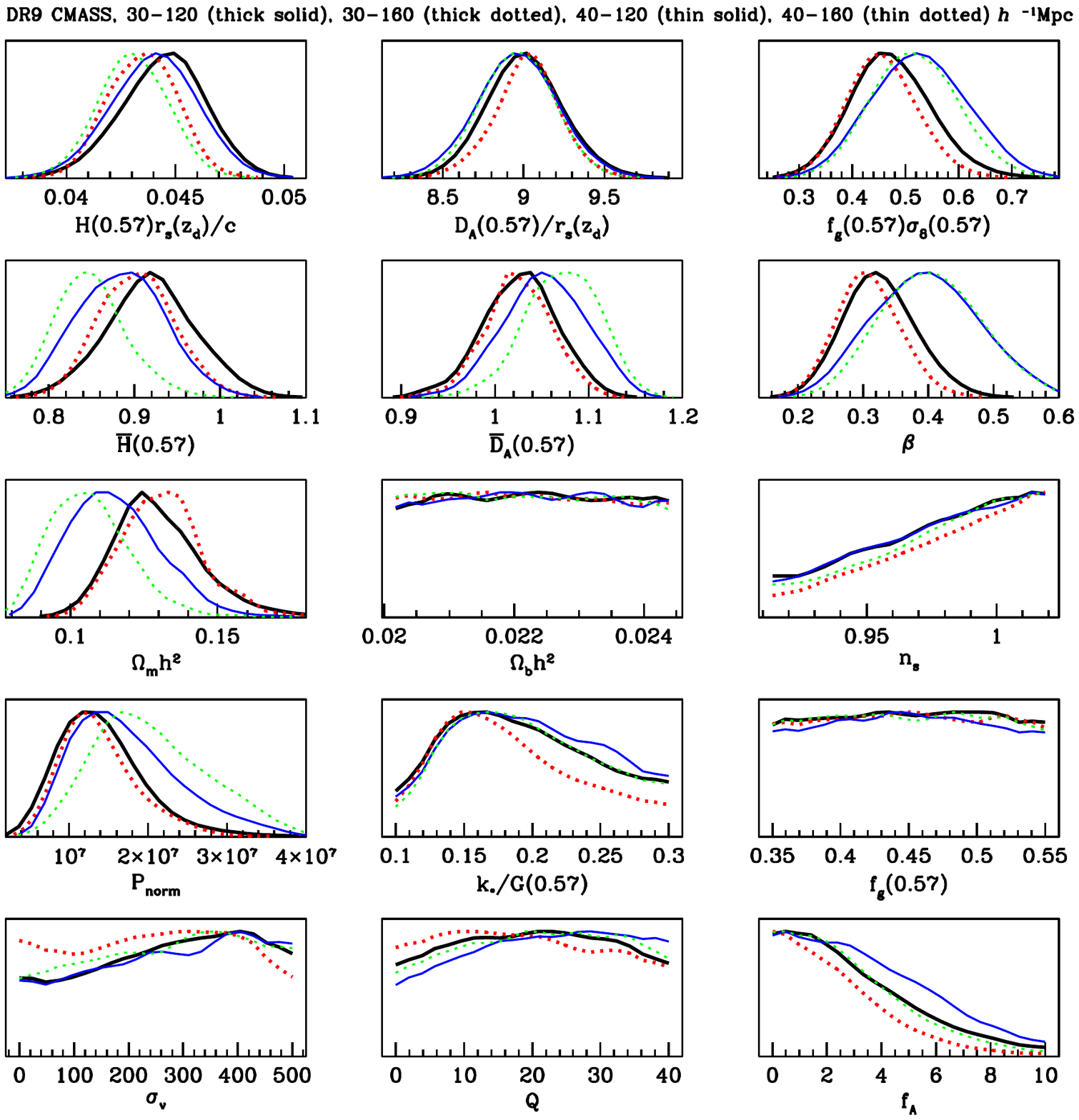}
\caption{The 1D marginalized probability distribution of parameters 
estimated from DR9 BOSS North and South CMASS samples. The different line
types denote different scale ranges used in our analysis:
30-120$\,h^{-1}$Mpc (thick solid); 30-160$\,h^{-1}$Mpc (thick dotted);
40-120$\,h^{-1}$Mpc (thin solid); 40-160$\,h^{-1}$Mpc (thin dotted).}
\label{fig:params_pdf}
\end{figure}

\begin{figure}
\centering
\includegraphics[width=\columnwidth,clip]{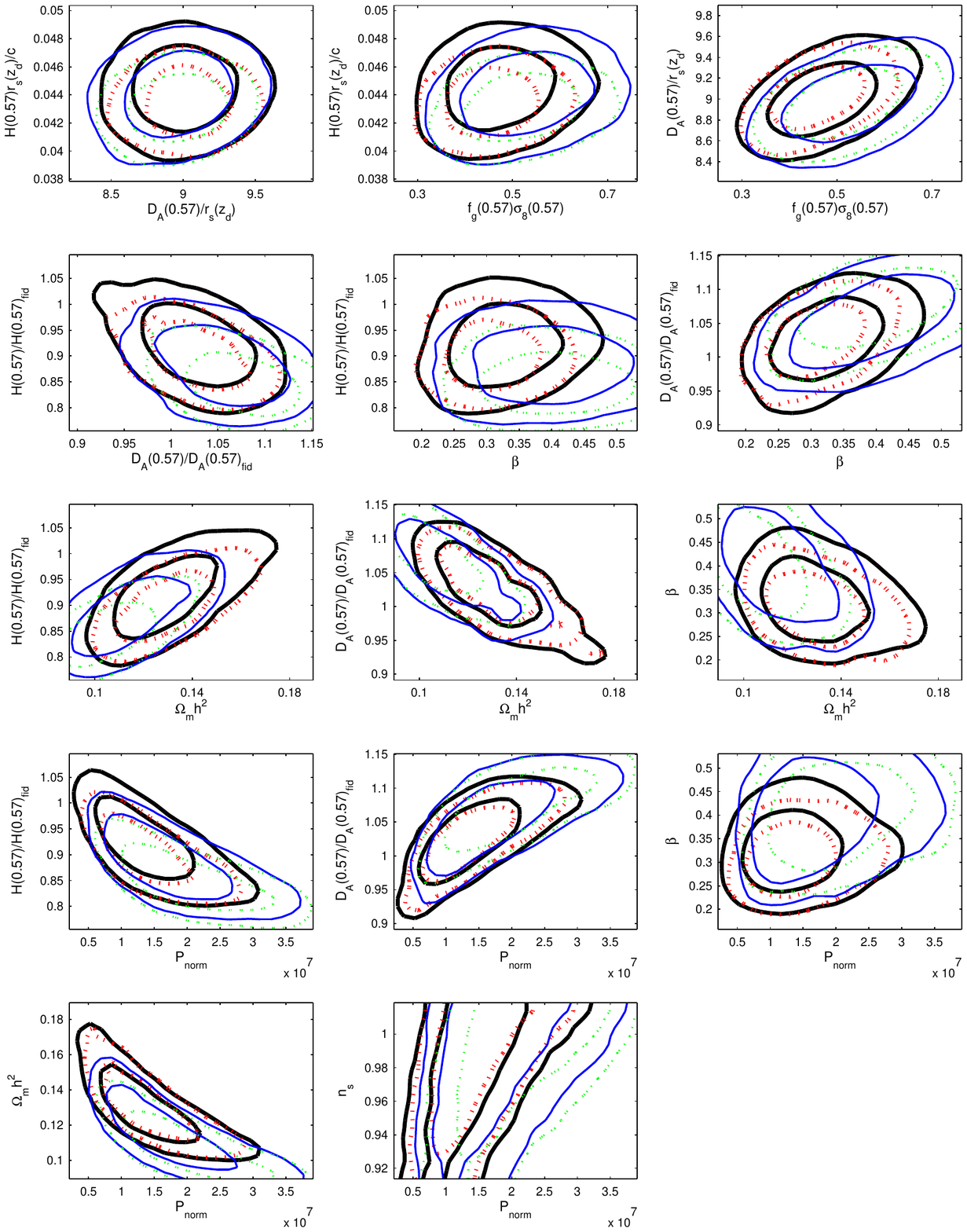}
\caption{The 2D joint confidence contours (68\% and 95\%) of the parameters
estimated from DR9 BOSS North and South CMASS samples, with the same line
types as in Fig.\ref{fig:params_pdf}. Only the key parameters and
parameters with significant correlations are shown.}
\label{fig:params_2D}
\end{figure}
Clearly, the data on larger scales, 120-160$\,h^{-1}$Mpc, do not
add significant amount of information to the parameter estimation.
This is because data on the scale range of 120-160$\,h^{-1}$Mpc
is very noisy (see Fig.\ref{fig:xi2d}). 
Table \ref{table:chi2} lists the $\chi^2_{min}$ for the scale ranges that we have considered.
Note that the number of fitted parameters is 12.
The addition of data on the scale range of 120-160$\,h^{-1}$Mpc
significantly increases the $\chi^2$ per degree of freedom.
\begin{table*}
\begin{center}
\begin{tabular}{lccc}\hline
scale range & $N_{data}$ & $\chi^2_{min}$ & $\chi^2_{min,pdf}$\\ \hline
30-120$\,h^{-1}$Mpc & 208    & 225.97   & 1.15\\
30-160$\,h^{-1}$Mpc & 390   & 529.17  & 1.40 \\
40-120$\,h^{-1}$Mpc & 198    & 214.29  & 1.15 \\
40-160$\,h^{-1}$Mpc & 380  & 511.97 & 1.39\\
\hline
\end{tabular}
\end{center}
\caption{The $\chi^2_{min}$ for the four different scale ranges of the SDSS DR9 BOSS CMASS samples.
Note that the number of fitted parameters is 12.} 
\label{table:chi2}
\end{table*}

On the other hand, the inclusion of data on the scale of
30-40$\,h^{-1}$Mpc is critical to placing a tight constraint
on $\beta$ (see Fig.\ref{fig:params_pdf}); this is as expected, since more information on
RSD comes from smaller scales, where the measured 2DCF is
smooth. Tighter constraints on $\beta$ leads to tighter constraints
on $H(0.57)$, $D_A(z)$, and $f_g(0.57)\sigma_8(0.57)$
due to parameter correlations (see Fig.{\ref{fig:params_pdf}}).
We do not use the scale range below 30$\,h^{-1}$Mpc, where
our current model is not expected to fit well as we have not
included modeling for the mixing of nonlinearity and RSD
on the smallest scales.

We choose the results from scale range of $30-120\,h^{-1}$Mpc as our fiducial results,
as they are less affected by noisy data, and retain the information on RSD.
Tables \ref{table:means} and \ref{table:covmat} give the marginalized means 
and the normalized covariance matrix for the key parameters that we have measured
or derived from measurements.
Table \ref{table:means} also includes results from using the scale range 
of $30-160\,h^{-1}$Mpc for comparison.

\begin{table}
\begin{center}
\begin{tabular}{cll}\hline
&$30<s<120$ &$30<s<160$ \\ \hline
$	H(0.57)	$& 86.14 $\pm$ 4.77 & 84.52$\pm$4.04\\
$	D_A(0.57)$& 1396.11 $\pm$ 53.17 & 1394.04$\pm$47.64\\
$	\Omega_m h^2$& 0.130 $\pm$ 0.015&  0.131$\pm$  0.014\\
$	\beta	$&  0.326 $\pm$  0.055 &  0.309$\pm$  0.048\\
\hline
$	H(0.57) \,r_s(z_d)/c	$& 0.0444$\pm$0.0019 & 0.0435 $\pm$  0.015\\ 
$	D_A(0.57)/r_s(z_d)	$& 9.01 $\pm$ 0.23& 9.02$\pm$  0.20\\
$	f(0.57)\,\sigma_8(0.57)	$& 0.474 $\pm$  0.075 & 0.457$\pm$  0.065\\
\hline
\end{tabular}
\end{center}
\caption{
The mean and standard deviation of 
$\{H(0.57)$, $D_A(0.57)$, $\Omega_m h^2$, $\beta$,  $H(0.57) \,r_s(z_d)/c$, $D_A(0.57)/r_s(z_d)$, 
$f(0.57)\,\sigma_8(0.57)\}$ from SDSS DR9 CMASS samples, for the scale ranges $30<s<120h^{-1}$Mpc and $30<s<160h^{-1}$Mpc.
The unit of $H$ is $\Hunit$. The unit of $D_A$ is $\rm Mpc$.
} \label{table:means}
\end{table}

\begin{table*}
\begin{center} 
\begin{tabular}{crrrrrrrr}\hline
       &$H(0.57)$ &$D_A(0.57)$   &$\Omega_mh^2$ &$\beta$  &$H(0.57) \,r_s(z_d)/c$ &$D_A(0.57)/r_s(z_d)$ &$f(0.57)\sigma_8(0.57)$ \\ \hline
$H(0.57)$&          1.0000  &   -0.4659  &    0.6794  &    0.1742  &    0.8535  &    0.0432 &     0.0102\\
$D_A(0.57)$&       -0.4659 &     1.0000  &   -0.7592  &    0.4076  &   -0.0980  &    0.6535  &    0.5375\\
$\Omega_mh^2$&      0.6794 &    -0.7592  &    1.0000  &   -0.2659  &    0.2189  &   -0.0419  &   -0.2451\\
$\beta$&            0.1742   &   0.4076   &  -0.2659 &     1.0000  &    0.4056  &    0.3158  &    0.9008\\
$H(0.57) \,r_s(z_d)/c$& 0.8535  &   -0.0980&    0.2189 &     0.4056 &     1.0000 &     0.0403 &     0.1771\\
$D_A(0.57)/r_s(z_d)$& 0.0432     & 0.6535   &  -0.0419 &     0.3158 &     0.0403 &     1.0000  &    0.5333\\
$f(0.57)\sigma_8(0.57)$& 0.0102  &    0.5375 &    -0.2451&    0.9008 &     0.1771 &     0.5333  &    1.0000\\

\hline
\end{tabular}
\end{center}
\caption{Normalized covariance matrix of the measured and derived parameters, $\{H(0.57)$, $D_A(0.57)$, $\Omega_m h^2$, $\beta,$, 
$H(0.57) \,r_s(z_d)/c$, $D_A(0.57)/r_s(z_d)$, $f(0.57)\sigma_8(0.57)\}$ from SDSS DR9 CMASS samples for the scale ranges $30<s<120h^{-1}$Mpc.}
 \label{table:covmat}
\end{table*}

\section{Summary and Discussion}
\label{sec:conclusion}

We have analyzed the anisotropic two dimensional galaxy correlation function (2DCF) of the CMASS galaxy samples
from the Sloan Digital Sky Survey Data Release 9 (DR9) of the Baryon Oscillation Spectroscopic Survey 
(BOSS) data, and derived robust constraints on $H(0.57)r_s(z_d)/c$, $D_A(0.57)/r_s(z_d)$, and
$f_g(0.57)\sigma_8(0.57)$ (see Table \ref{table:means} and Table \ref{table:covmat}). 
While consistent with previous results using
the same data, our results have significantly better precision since we are using all the
information from the 2DCF in the scale range of interest.
Since our measurements have been marginalized over sufficiently wide priors for
the relevant parameters; they can be combined with other data to probe dark energy and gravity.

We found that the data beyond the scale of 120 $h^{-1}$Mpc are dominated by noise (see Table \ref{table:chi2}).
On the other hand, the inclusion of data below the scale of 40 $h^{-1}$Mpc is important for
constraining the redshift-space distortion parameter, and hence of the growth rate (see Fig.\ref{fig:params_pdf}).
We have chosen the scale range of 30 to 120 $h^{-1}$Mpc (the quasilinear and linear
scales at $z=0.57$) in obtaining our fiducial results.

We do not use the scale range below 30$\,h^{-1}$Mpc, where
our current model is not expected to fit well as we have not
included modeling for the mixing of nonlinearity and RSD
on the smallest scales. The more advanced modeling that
would apply to clustering on smaller scales cannot be validated
using the BOSS DR9 mocks, as these were produced 
using a second-order Lagrangian perturbation theory (2LPT) method, and
were calibrated to reproduce the clustering measurements between 30 
and 80$\,h^{-1}$Mpc \citep{Manera13}. 
Fig.1 shows that our model fits the data and the mocks well in the scale range used (indicated by the gray
band), and for transverse separations greater than 10 $\,h^{-1}$Mpc for the mocks. We carried out
MCMC runs with and without making the transverse cut at $\sigma\geq 10 \,h^{-1}$Mpc, and found that they give
very similar results. This may be due to the fact that the cut would only remove 
a relatively small number of data points (our bin size is 10$\,h^{-1}$Mpc x 10$\,h^{-1}$Mpc).

Note that the results from the BOSS DR9 CMASS north and south samples have significantly
smaller measurement uncertainties compared to the expectation based on the distribution of the
results from the mocks (compare Fig.\ref{fig:params_pdf} and Fig.\ref{fig:meanxhxd}). This may be due to the fact that
the measured 2DCF appears less noisy that those measured from the mocks (see Fig.\ref{fig:xi2d}),
which reflects a statistical property of the data.

\cite{Linder14} presented another independent analysis of DR9 BOSS data, using
the method from \cite{Song14}, which is a similar approach with a different theoretical model.
The main difference in methodology is that \cite{Linder14} effectively fixed the shape of $P(k)$, while we
marginalize over the shape of $P(k)$ by marginalizing over $\Omega_m h^2$, $\Omega_b h^2$, and $n_s$.
Our results are broadly similar to that of \cite{Linder14}, with 
the main difference being that \cite{Linder14} measured $H(0.57)$, $D_A(0.57)$, and $G_{\Theta}$ 
with either WMAP9 or Planck priors, while our measurements are independent of the CMB priors.

Our measurements of $H(0.57)$ and $D_A(0.57)$ are consistent with the expected values from
WMAP9 \citep{Bennett13} at 68\% confidence level. \cite{Spergel13} showed that Planck results \citep{PlanckXVI}
may be sensitive to systematic effects; they found that the difference between Planck and
WMAP 9 results are significantly reduced once the Planck data are cleaned in a consistent
and systematic manner.

The latest BOSS data (DR11) seem to give much more stringent results than DR9 (see
\cite{Anderson13,Chuang13,Samushia13,Sanchez13}). It will be interesting to apply our method
to BOSS DR11 data, once they are publicly available. 
As sufficiently large mock catalogues become available, we will be able to
further validate our methodology for application to the Euclid GC data.

\section*{Acknowledgments}

Computational facilities for this project were provided by the OU Supercomputing Center 
for Education and Research (OSCER) at the University of Oklahoma (OU).
I am grateful to OSCER Director Henry Neeman for invaluable technical support,
and to Chia-Hsun Chuang for help using the public BOSS DR9 data and for sharing the 2DCF 
computed from the 600 mocks.
This work was supported in part by DOE grant DE-SC0009956, and NASA grant 12-EUCLID12-0004.

Funding for SDSS-III has been provided by the Alfred P. Sloan Foundation, the Participating Institutions, the National Science Foundation, and the U.S. Department of Energy Office of Science. The SDSS-III web site is http://www.sdss3.org/.

SDSS-III is managed by the Astrophysical Research Consortium for the Participating Institutions of the SDSS-III Collaboration including the University of Arizona, the Brazilian Participation Group, Brookhaven National Laboratory, Carnegie Mellon University, University of Florida, the French Participation Group, the German Participation Group, Harvard University, the Instituto de Astrofisica de Canarias, the Michigan State/Notre Dame/JINA Participation Group, Johns Hopkins University, Lawrence Berkeley National Laboratory, Max Planck Institute for Astrophysics, Max Planck Institute for Extraterrestrial Physics, New Mexico State University, New York University, Ohio State University, Pennsylvania State University, University of Portsmouth, Princeton University, the Spanish Participation Group, University of Tokyo, University of Utah, Vanderbilt University, University of Virginia, University of Washington, and Yale University.


\setlength{\bibhang}{2.0em}

\label{lastpage}

\end{document}